\documentclass[notitlepage,a4,10.5pt]{article}%,twocolumn]{revtex4}

\usepackage{graphicx}

\usepackage{amsmath}%
\usepackage{amsfonts}%
\usepackage{amssymb}%
\usepackage{mathrsfs}
\usepackage{amsthm}
\usepackage{float}

\usepackage{authblk}

\usepackage[left=2.5cm,right=2.5cm,top=3cm,bottom=3cm]{geometry} 

\usepackage{xcolor}

\newcommand{\mc}{\mathcal}
\newcommand{\cp}{\times}

\newcommand{\bol}{\boldsymbol}

\newcommand{\abs}[1]{\left\lvert{#1}\right\rvert}

\newcommand{\lr}[1]{\left({#1}\right)}
\newcommand{\lrs}[1]{\left[{#1}\right]}
\newcommand{\lrc}[1]{\left\{{#1}\right\}}

\newcommand{\p}{\partial}

\newcommand{\ti}[1]{\textit{#1}}
\newcommand{\tb}[1]{\textbf{#1}}

\begin{document}

\title{%Guiding Center Correction to the Fluid Momentum Equation\\ Leading to 
Guiding Center Derivation of the Generalized Hasegawa-Mima\\ Equation for Drift Wave Turbulence in Curved Magnetic Fields}
\author[1]{Naoki Sato} 
\author[2]{Michio Yamada}
\affil[1]{Graduate School of Frontier Sciences, \protect\\ The University of Tokyo, Kashiwa, Chiba 277-8561, Japan \protect\\ Email: sato\_naoki@edu.k.u-tokyo.ac.jp}
%\affil[2]{Graduate School of Frontier Sciences, \protect\\ The University of Tokyo, Kashiwa, Chiba 277-8561, Japan
%\protect \\ Email: saito@ppl.k.u-tokyo.ac.jp}
\affil[2]{Research Institute for Mathematical Sciences, \protect\\ Kyoto University, Kyoto 606-8502, Japan
\protect \\ Email: yamada@kurims.kyoto-u.ac.jp}
\date{\today}
\setcounter{Maxaffil}{0}
\renewcommand\Affilfont{\itshape\small}

%\twocolumn[
 % \begin{@twocolumnfalse}
    \maketitle
    \begin{abstract}
    Recently, a generalized Hasegawa-Mima (gHM) equation describing drift wave turbulence in curved magnetic fields has been derived in [N. Sato and M. Yamada, J. Plasma Phys. (2022), vol. 88, 905880319] for an ion-electron plasma modeled as a two-fluid system. 
    In this work, we show that a mathematically equivalent GHM equation can be obtained within the kinetic framework of guiding center motion, 
    and that the relevant drift wave turbulence ordering can be further relaxed, effectively generalizing the applicability of the equation to any magnetic field geometry and electron spatial density, in the sense that no ordering requirements involve spatial derivatives of the magnetic field or the electron spatial density. 
    %This result stems from a guiding center correction to the %usual 
    %which represents %the existence of 
    %an effective potential energy associated with the kinetic energy of ExB drift motion, which is a pure spatial function. 
   % The noncanonical Hamiltonian structure of the equation is discussed, and conditions for the nonlinear stability of steady solutions are derived through the energy-Casimir stability criterion. Finally, these results are applied to describe drift waves and infer the existence of stable toroidal zonal flows with radial shear in dipole magnetic fields.
   %The noncanonical Hamiltonian structure of the equation is discussed, and conditions for the nonlinear stability of steady solutions are derived through the energy-Casimir stability criterion. Finally, these results are applied to describe drift waves and infer the existence of stable toroidal zonal flows with radial shear in dipole magnetic fields.
    \end{abstract}
%\vspace{5mm}
%\end{@twocolumnfalse}
%  ]

\section{Introduction}
%1. evaluate everything at x
%2. cannot estimate polarization drift from charged partcicle EoM
%3. need to evaluate $\rho_s/L_{\perp}$
The Hasegawa-Mima (HM) equation \cite{HM,HM2} 
is a nonlinear equation describing the turbulent behavior of  
electric potential and spatial density in a quasi-neutral plasma,   
made of hot thermalized electrons and cold ions,  
permeated by a strong, straight, and homogeneous magnetic field,  
and evolving over time scales long compared with the period of cyclotron motion. 
The nonlinearity of the HM equation is caused by the convection of the $\bol{E}\cp\bol{B}$ velocity associated with the polarization drift. 
In the presence of an electron density gradient, 
solutions of the linearized HM equation are the characteristic drift waves, whose interaction gives rise to drift wave turbulence. 
The HM equation shares the same mathematical structure with 
the quasi-geostrophic equation characterizing  
atmospheric motion over the surface of rotating planets 
\cite{Charney,Charney3} due to the similarity between the Lorentz force and the Coriolis force, and contains 2-dimensional incompressible vorticity dynamics as a special case \cite{Horton}.
Despite its relative simplicity, the physical significance of the HM equation stems from its ability to capture  
essential features of 2-dimensional plasma and fluid turbulence \cite{Batchelor},  
including onset of inverse turbulent cascades of energy \cite{Kraichnan2,Kraichnan,Rivera,Xiao} 
and self-organization of large scale structures and zonal flows \cite{HM5,Hasegawa85,Horton2,Fujisawa,Diamond}.  

One of the key assumptions behind the HM equation is that both the background magnetic field and the electron spatial density change over a spatial scale $L$ that is large compared to the typical turbulence wavelength across the magnetic field $k^{-1}_{\perp}$, i.e. $k_{\perp}L>>1$.  
This hypothesis effectively restricts the applicability of the HM equation to 
plasmas with a small density gradient and to magnetic fields with small curvature or  field inhomogeneities.  
However, experimental observations pertaining to plasmas confined by dipole magnetic fields \cite{Boxer,Kenmochi} suggest the existence of drift wave turbulence and zonal flows in systems where 
both the electron spatial density and the magnetic field are characterized by strong gradients over spatial scales comparable to that of electric field and density fluctuations (these low frequency fluctuations are often referred to as entropy modes \cite{Garnier}). 
In principle, an accurate description of electromagnetic turbulence in such setting could be obtained with the aid of nonlinear gyrokinetic theory \cite{Hahm96a,Hahm96b,Hahm09}. 
Nevertheless, it is natural to ask whether the HM equation can be generalized 
to allow strong magnetic field and density inhomogeneities while 
maintaining a single governing equation for the electric potential $\varphi$. 
%could generalize the formalism to hot ions by including drifts and keeping track of $B^{\ast}$, but how to change the energy??
%Point of discussion $\nabla\cdot\bol{v}_{\bol{E}}$ and nonHailtonian origin of the ordering difficulty. 
In \cite{gHM} this question has been answered positively, and a generalized Hasegawa-Mima (gHM) equation has been obtained from a two-fluid model \cite{Haz98} of an ion-electron plasma in the form below (see section 2 for the definition of the physical quantities appearing in the equation):
\begin{equation}
\frac{\p}{\p t}\lrs{\lambda A_e\varphi-\sigma\nabla\cdot\lr{A_e\frac{\nabla_{\perp}\varphi}{B^2}}}=\nabla\cdot\lrs{A_e\lr{\sigma\frac{\bol{B}\cdot\nabla\cp\bol{v}_{\bol{E}}}{B^2}-1}\bol{v}_{\bol{E}}}\label{eq1}.
\end{equation}
In particular, the ordering used in \cite{gHM}  
to derive the gHM equation \eqref{eq1} only involves one ordering condition on the 
spatial derivatives of the magnetic field and the electron spatial density,
effectively extending the range of the HM equation to general magnetic field geometries (see section 4 for details). 

The aim of this paper is twofold. First, we want to show that
a guiding center generalized Hasegawa-Mima (GHM) equation \eqref{GHMgc5}, 
mathematically equivalent to the gHM equation \eqref{eq1}, can be derived within the framework of guiding center dynamics \cite{Cary,Northrop} (here, the uppercase letter G is used to emphasize that the GHM equation can be derived under weaker assumptions, and thus it is more general than the gHM equation). 
Secondly, we wish to show that the guiding center drift wave turbulence ordering required to derive the GHM equation does not involve conditions on the spatial derivatives of the magnetic field or the electron spatial density, and it is therefore weaker than the two-fluid drift wave turbulence ordering used to derive the gHM equation from a two-fluid model in \cite{gHM}. 
This difference does not originate from a discrepancy between two-fluid theory and guiding center theory. In fact, we will see that the new GHM equation can be obtained from a two-fluid ordering equivalent to the guiding center drift wave turbulence ordering by adding tailored higher-order terms to the gHM equation.  %discrepancy originates from the fact that, upon integration of the 
%evolution equation for the ion distribution function, guiding center theory predicts an additional effective force on the right-hand side of the fluid momentum equation. 
%This correction arises from the kinetic energy of $\bol{E}\cp\bol{B}$ dynamics, 
%which is a pure spatial function and thus behaves
%as an effective potential energy with an associated guiding center drift. 
%Secondly, we wish to elucidate the Hamiltonian structure \cite{PJM} of the gHM equation \eqref{eq1}, 
%and use it to infer the stability properties of steady solutions. 
%Finally, we want to determine whether zonal flows can form in dipole magnetic fields, 
%and characterize drift waves in dipole geometry.
%The aim of these paper is threefold. 
%First, we want to show that
%the gHM equation \eqref{eq1} can be derived within the framework of  
%guiding center dynamics \cite{Cary}.  
%Thirdly, we wish to elucidate the Hamiltonian structure \cite{PJM} of the GHM equation, 
%and use it to infer the stability properties of steady solutions. 
%In particular, we want to determine whether zonal flows can form in dipole magnetic fields, 
%and characterize drift waves in dipole geometry.

The present paper is organized as follows.
In section 2, we derive the GHM equation by considering the evolution of the phase space distribution function of a magnetized plasma according to the guiding center equations of motion. 
In section 3, we discuss the constants of motion of the GHM equation, and obtain a more general form of the generalized enstrophy presented in \cite{gHM}.
In section 4 we examine the relationship between the ordering used to derive the gHM equation \eqref{eq1} from a two-fluid model in \cite{gHM} and the ordering used here to obtain the GHM equation \eqref{GHMgc5} from guiding center theory. In particular, we show that the guiding-center ordering does not involve any conditions on the spatial derivatives of the magnetic field or the electron spatial density. 
%In section 5 we examine the algebraic structure of the gHM equation, and obtain sufficient conditions on magnetic field and electron spatial density under which the GHM equation  defines a noncanonical Hamiltonian system. 
%These results are consistent with the Hamiltonian structure of the standard HM equation 
%\cite{Weinstein,Tassi,Hazeltine,Hazeltine2}.  
%In section 6 we prove a theorem concerning the nonlinear stability of steady solutions of the GHM equation by applying the energy-Casimir method \cite{Holm,Tronci,Rein}. This result generalizes Arnold's stability criterion for a 2-dimensional fluid flow \cite{Arnold}. 
%In section 7 we show that stable toroidal zonal flows can form in dipole magnetic fields, 
%and characterize the angular frequency of drift waves in dipole geometry.
Concluding remarks are given in section 5.

\section{Derivation of the GHM equation within the kinetic framework of guiding center motion}

In this section, we derive the GHM equation for an ion-electron plasma obeying the guiding-center equations of motion under an appropriate drift wave turbulence ordering. 

\subsection{Guiding center ordering}

We consider a guiding-center plasma made of ions and electrons within a region $\Omega\subseteq\mathbb{R}^3$ permeated by a static magnetic field $\bol{B}\lr{\bol{x}}\neq\bol{0}$ with modulus $B$, and where $\bol{x}=\lr{x,y,z}\in\mathbb{R}^3$ are Cartesian coordinates. Let $\bol{E}=-\nabla \varphi$ denote the electric field, with $\varphi\lr{\bol{x},t}$ the electric potential ($t$ is the time variable), $\bol{E}_{\perp}$ the component of $\bol{E}$ perpendicular to $\bol{B}$, and $\bol{E}_{\parallel}$ the component of $\bol{E}$ parallel to $\bol{B}$. 
The small ordering parameter for the guiding-center expansion will be denoted by $\epsilon>0$, the spatial scale of the system by $L$, the time scale of the system by $\tau$, the ion gyroradius by $\bol{\rho}$, and the ion cyclotron frequency by $\omega_{c}=ZeB/m$ where $Z\in\mathbb{N}$ and $Ze$ and $m$ are the ion electric charge and mass respectively. 
Let $\bol{X}=\bol{x}-\bol{\rho}$ denote the ion guiding center position,  $\bol{b}=\bol{B}/B$ the unit vector along $\bol{B}$,   $\mu=m\lr{\bol{{\bol{v}}_{\perp}-\bol{v}_{\bol{E}}}}^2/2B$ the lowest order magnetic moment, $\bol{v}=\dot{\bol{x}}$ the charged particle velocity, 
${\bol{v}}_{\perp}=\bol{b}\times\lr{{\bol{v}}\times\bol{b}}$ the charged particle velocity perpendicular to $\bol{B}$,  and
\begin{equation}
\bol{v}_{\bol{E}}=\frac{\bol{E}\times\bol{B}}{B^2}, 
\end{equation}
the $\bol{E}\times\bol{B}$ velocity. Then, the ion guiding center equations of motion obtained from the Northrop phase space guiding center Lagrangian \cite{Cary,Northrop} are  
\begin{subequations}
\begin{align}
&m\dot{u}\bol{b}=Ze\lr{\bol{E}'+\dot{\bol{X}}\cp\bol{B}'},\\
&\dot{\mu}=0,\\
&\dot{\zeta}=\omega_c,
\end{align}\label{GCEoM}
\end{subequations}
which can be equivalently written as
\begin{subequations}
\begin{align}
&\dot{\bol{X}}=u\frac{\bol{B}'}{B_{\parallel}'}+\bol{E}'\times\frac{\bol{b}}{B_{\parallel}'},\label{dotX}\\
&\dot{u}=\frac{Ze}{m}\frac{\bol{B}'\cdot\bol{E}'}{B'_{\parallel}},\label{dotu}\\
&\dot{\mu}=0,\\
&\dot\zeta=\omega_c,
\end{align}
\end{subequations}
where $\zeta$ is the gyrophase and 
\begin{subequations}
\begin{align}
&u=\dot{\bol{X}}\cdot\bol{b},\\&
Ze\varphi'=Ze\varphi+\mu B+\frac{m}{2}\bol{v}_{\bol{E}}^2,\\
&\bol{A}'=\bol{A}+\frac{m}{Ze}\lr{u\bol{b}+\bol{v}_{\bol{E}}}
\\&\bol{E}'=-\nabla\varphi'-\frac{\p\bol{A}'}{\p t},\\&\bol{B}=\nabla\cp\bol{A},~~~~\bol{B}'=\nabla\cp\bol{A}',~~~~B_{\parallel}'=\bol{B}'\cdot\bol{b}.\label{u}
\end{align}
\end{subequations}
Here, we observe that $u$ represents the component of the guiding center velocity parallel to $\bol{B}$. Furthermore, notice that
the following guiding center drift velocities 
$\bol{v}_{\bol{E}}$ ($\bol{E}\times\bol{B}$ drift), $\bol{v}_{\nabla}$ ($\nabla B$ drift), and $\bol{v}_{\kappa}$ (curvature drift) are contained in the right-hand side of \eqref{dotX} according to
\begin{equation}
\bol{v}_{\bol{E}}'=\frac{\bol{b}\times\nabla\varphi}{B'_{\parallel}},~~~~\bol{v}_{\nabla}'=\frac{\mu}{Ze}\frac{\bol{b}\times\nabla B}{B'_{\parallel}},~~~~\bol{v}_{\kappa}'=\frac{mu^2}{ZeB_{\parallel}'}\nabla\times\bol{b},
\end{equation}
where the $'$ symbol is used to emphasize that the correction $B_{\parallel}'$
of the magnetic field $B$ caused by the term $m\nabla\times\lr{u\bol{b}+\bol{v}_{\bol{E}}}/Ze$ in $\bol{B}'$ is used in these formulas.
Similarly, the polarization drift $\bol{v}_{\rm pol}$ is  included in \eqref{dotX} according to
\begin{equation}
\bol{v}_{\rm pol}'=\frac{m}{Ze}\frac{\bol{b}\cp\frac{\p\bol{v}_{\bol{E}}}{\p t}}{B_{\parallel}'}+\bol{v}_{\bol{E}}'-\bol{v}_{\bol{E}}+\frac{m}{2Ze}\frac{\bol{b}\cp\nabla\bol{v}_{\bol{E}}^2}{B_{\parallel}'}.
\end{equation}
The physical meaning carried by this  expression will become clear later.
Equation \eqref{dotX} also includes a  further drift term  
\begin{equation}
\bol{v}_{\ast}'=\frac{mu}{ZeB_{\parallel}'}{\nabla\cp{\bol{v}_{\bol{E}}}},
\end{equation}
which originates from 
the effective magnetic field  $\frac{m}{Ze}\nabla\cp{\bol{v}_{\bol{E}}}$ 
%and the effective potential energy $m\bol{v}_{\bol{E}}^2/2$ 
associated with  $\bol{E}\cp\bol{B}$ motion.
The total guiding center velocity can thus be written as
\begin{equation}
\dot{\bol{X}}=u\frac{\bol{B}}{B_{\parallel}'}+\bol{v}_{\bol{E}}+\bol{v}'_{\nabla}+\bol{v}_{\kappa}'+\bol{v}_{\rm pol}'+\bol{v}'_{\ast}.
\end{equation}
%$\bol{\rho}$ 
%has been divided into a gyrophase independent term $\bar{\bol{\rho}}$ and a gyrophase dependent term $\tilde{\bol{\rho}}$, according to  $\bol{\rho}=\bar{\bol{\rho}}+\tilde{\bol{\rho}}$, 
For completeness we recall that 
%this expression coincides with the  
%and 
the gyroradius, which defines the coordinate transformation $\bol{X}=\bol{x}-\bol{\rho}$, 
is an oscillatory (gyrophse dependent) term given by 
\begin{equation}
\bol{\rho}=\frac{m}{Ze}\frac{\bol{b}\cp\lr{{\bol{v}}-\bol{v}_{\bol{E}}}}{B^2}
%\bar{\bol{\rho}}+\tilde{\bol{\rho}},~~~~\bar{\bol{\rho}}=\frac{m}{Ze}\frac{\bol{B}\cp\bol{v}_{\bol{E}}}{B^2}=\frac{\bol{E}_{\perp}}{\omega_c B},~~~~\tilde{\bol{\rho}}=\frac{m}{Ze}\frac{\bol{B}\cp\lr{\dot{\bol{x}}-\bol{v}_{\bol{E}}}}{B^2}
.\label{gyro}
\end{equation}
This term is removed from the Northrop guiding center phase space Lagrangian density $\mc{L}_{Ngc}$ 
%$\mc{L}_{gc}$ 
by appropriate subtraction of total time derivatives (on this point, see section III.C and appendix A of  \cite{Cary}).  
%can been neglected under the assumption that it vanishes when averaged over a gyroperiod. 
%the gyrophase dependent term $\tilde{\bol{\rho}}$ is neglected. 
%In the following, we shall assume $\tilde{\bol{\rho}}$ to be an oscillatory term that vanishes when averaged with respect to the gyrophase $\zeta$, and set $\bol{\rho}=\bar{\bol{\rho}}$. 
Here, the Northrop guiding center phase space Lagrangian density $\mc{L}_{Ngc}$ is obtained by
expansion of the charged particle phase space Lagrangian density $\mc{L}=\mc{L}_{Ngc}+o\lr{\epsilon}$  
according to the guiding center ordering. 
Since the Northrop guiding center phase space Lagrangian density $\mc{L}_{Ngc}$ is independent of ${\bol{\rho}}$, it is also independent of the gyrophase $\zeta$, 
leading to conservation of the conjugate momentum $\mu$ by the Noether theorem.   
The guiding-center ordering required for the conservation of the magnetic moment $\mu$ is given in table I of \cite{Cary}, which we report in table \ref{tab1}.
This ordering represents the starting point that we will use to
construct a more restrictive ordering leading to the GHM equation.
We conclude by observing that the ordering parameter $\epsilon$ arises from the
physical constant 
\begin{equation}
\sigma=\frac{m}{Ze}<<1,
\end{equation}
which is small for elementary particles such as ions and electrons. 
The constant $\sigma^{-1}$ always multiplies the electromagnetic fields within the
charged particle phase space Lagrangian density $\mc{L}\lr{\bol{x},{\bol{\bol{v}}},t}=\lr{\sigma^{-1}\bol{A}+{\bol{\bol{v}}}}\cdot\dot{\bol{x}}-\frac{1}{2}{\bol{\bol{v}}}^2-\sigma^{-1}\varphi$, which is the reason why $\bol{E}$ and $\bol{B}$ are treated as large fields in table \ref{tab1}. 
%In this notation, $\bol{A}$ is the vector potential such that $\bol{B}=\nabla\cp\bol{A}$. 

\begin{table}
\begin{center}
\begin{tabular}{c c c c c c} 
 \hline
 \hline
 Order & Dimensionless & Fields & Distances & Rates & Velocities\\  
 \hline 
 $\epsilon^{-1}$ & & $\bol{B},\bol{E}_{\perp}$& & $\omega_{c}$ &\\ 
 $1$ &  & $\bol{E}_{\parallel}$ & $L$ & $\bol{v}/L,\bol{v}_{\bol{E}}/L,\tau^{-1}$ &$\bol{v},\bol{v}_{\bol{E}}$\\
 $\epsilon$ & $\rho/L$, $\lr{\omega_{c}\tau}^{-1}$ &  & $\rho$ & $\bol{v}_{\nabla}/L,\bol{v}_{\kappa}/L,\bol{v}_{\rm pol}/L$  & $\bol{v}_{\nabla},\bol{v}_{\kappa},\bol{v}_{\rm pol}$\\
 \hline
 \hline
\end{tabular}
\caption{\label{tab1} Guiding center ordering required for the existence of the first adiabatic invariant $\mu$ (see \cite{Cary}).}
\end{center}
\end{table}

%\subsection{Continuity equation for the ion density}

\subsection{Derivation of the GHM equation from a drift wave turbulence ordering within guiding center theory}
We start by assuming that the guiding center ordering presented in table \ref{tab1} holds, and gradually impose additional conditions to obtain the relevant drift wave turbulence ordering. 
From now on we set $\tau=\tau_d$, with $\tau_d$ the drift turbulence time scale. Recall that $\tau_d\omega_c\sim\epsilon^{-1}$. 
%Let $f\lr{\bol{X},\mu,u,t}$ denote the ion guiding-center probability  density function (distribution function) defined with respect to the invariant measure (preserved phase space volume) associated with the guiding-center equations of motion,
%\begin{equation}
%d\Pi= B_{\parallel}^{\ast}\,d\mu dv_{\parallel}d\bol{X}.
%\end{equation}
%The proof that this volume element is invariant under the flow generated by guiding center dynamics can be found in \cite{Cary}. 
%In the absence of particle collisions, conservation of probability $fd\Pi$  implies that the distribution function $f$ obeys the Liouville equation
%\begin{equation}
%\frac{\p}{\p t}\lr{B_{\parallel}^{\ast}f}+\nabla\cdot\lr{B_{\parallel}^{\ast}\dot{\bol{X}}f}+\frac{\p}{\p u}\lr{B_{\parallel}^{\ast}\dot{u}f}=0,
%\end{equation}
%where $\nabla$ is the gradient operator with respect to the guiding center position $\bol{X}$, and we used the fact that $\dot{\mu}=0$. 
Let $f\lr{\bol{p},\bol{x},t}$ denote the ion distribution function in the  canonical phase space $\lr{\bol{p},\bol{x}}$ of charged particle dynamics.
%Assuming the plasma to be collisionless, 
The distribution function $f$ satisfies the Boltzmann equation  
\begin{equation}
\frac{\p f}{\p t}=-\frac{\p}{\p\bol{p}}\cdot\lr{\dot{\bol{p}}f}-\frac{\p }{\p\bol{x}}\cdot\lr{\dot{\bol{x}}f}+\lr{\frac{df}{dt}}_{\rm c},\label{Liouville}
\end{equation}
where the last term on the right-hand side describes particle collisions. 
Introducing the ion spatial density $n\lr{\bol{x},t}=\int_{\mathbb{R}^3}f\,d\bol{p}$, integrating equation \eqref{Liouville} with respect to the momentum variables, 
and assuming $\lim_{\abs{\bol{p}}\rightarrow \infty}f=0$, 
we obtain the ion continuity equation 
\begin{equation}
\frac{\p n}{\p t}=-\nabla\cdot\lr{\langle\dot{\bol{x}}\rangle n},\label{ni}
\end{equation}
where
\begin{equation}
\langle\dot{\bol{x}}\rangle=\frac{1}{n}\int_{\mathbb{R}^3}f\dot{\bol{x}}\,d\bol{p}=\frac{1}{n}\int_{\mathbb{R}^3}f\lr{\dot{\bol{X}}+\dot{\bol{\rho}}}\,d\bol{p}=\left\langle\dot{\bol{X}}\right\rangle,\label{avdotx}
\end{equation}
is the ensemble averaged ion velocity at a given position $\bol{x}=\bol{X}+{\bol{\rho}}$. Notice that in the last passage we used the fact that by hypothesis $\bol{\rho}$ is an oscillatory term such that the ensemble average $\langle\dot{\bol{\rho}}\rangle$ identically vanishes. 
We also remark that the collision term in \eqref{Liouville} vanishes upon integration in momentum space because we assume that collisions result in deflections in velocity space that do not change the local particle number.
%Let us now consider all contributions to the continuity equation \eqref{ni} that are greater than $\epsilon^2$. 
Now suppose that the %ion plasma is cold along the magnetic field  
%In particular, for the dynamics perpendicular to the magnetic field we assume that $k_BT_c\sim \mu B$ is small with $k_B$ the Boltzmann constant and $T_c$ a characteristic ion temperature associated with cyclotron dynamics. Similarly, 
parallel velocity $u$ is small (the time scale $\tau_b$ of dynamics along $\bol{B}$ is long):  
%The precise ordering conditions are the following: 
\begin{equation}
%\frac{k_BT_c}{\frac{m}{2}\bol{v}_{\bol{E}}^2}\sim 
u\frac{\tau_d}{L}\sim \frac{\tau_d}{\tau_b}\sim\epsilon^2.\label{usmall}
\end{equation}
Note that consistency with \eqref{dotu} requires that the component $E_{\parallel}'$ of $\bol{E}'$ along $\bol{B}'$ is small, i.e. $\bol{B}'\cdot\bol{E}'/B_{\parallel}'E_{\perp}\sim\epsilon^3$ (physically, this means that the electric field experienced by a charged particle along the magnetic field is negligible). 
%In particular, this also implies that $E_{\parallel}'/E_{\perp}'\sim \epsilon^3$. 
Let us now consider all contributions to the continuity equation \eqref{ni} that are greater than $\epsilon^2$.  
From the ordering condition  \eqref{usmall} it readily follows that  the only surviving terms in \eqref{dotX} are those involving the $\bol{E}\times\bol{B}$ drift velocity and the $\nabla B$ drift. In particular, observing that $B'_{\parallel}=B\lr{1+o\lr{\epsilon}}$, we have
\begin{equation}
\dot{\bol{X}}=\bol{v}_{\bol{E}}+\bol{v}'_{\nabla}+\bol{v}_{\rm pol}+o\lr{\epsilon^2},\label{ve1}
\end{equation}
where the polarization drift $\bol{v}_{\rm pol}$ now has expression
\begin{equation}
\bol{v}_{\rm pol}=\sigma\frac{\bol{b}\cp{\frac{d\bol{v}_{\bol{E}}}{dt}}}{B},~~~~\frac{d\bol{v}_{\bol{E}}}{dt}=\frac{\p\bol{v}_{\bol{E}}}{\p t}+\bol{v}_{\bol{E}}\cdot\nabla\bol{v}_{\bol{E}},\label{vpol1}
\end{equation}
and where we used the fact that
\begin{equation}
\bol{v}_{\bol{E}}+\bol{v}_{\rm pol}'=\bol{v}_{\bol{E}}-\sigma\frac{\bol{b}\cdot\nabla\cp\bol{v}_{\bol{E}}}{B}\bol{v}_{\bol{E}}+\sigma\frac{\bol{b}\cp\lr{\frac{\p\bol{v}_{\bol{E}}}{\p t}+\frac{1}{2}\nabla\bol{v}_{\bol{E}}^2}}{B}+o\lr{\epsilon^2}=\bol{v}_{\bol{E}}+\bol{v}_{\rm pol}+o\lr{\epsilon^2}.
\end{equation}
%at first order in $\epsilon$ equation \eqref{ft} thus reduces to
%\begin{equation}
%\frac{\p f}{\p t}+\frac{1}{B}\nabla\cdot\lr{B\bol{v}'_{\bol{E}}f}=0.
%\end{equation}
%Integrating this expression with respect to $\mu$ and $v_{\parallel}$, 
%and introducing the ion guiding-center spatial density $n\lr{\bol{X},t}=\int fB_{\parallel}^{\ast}d\mu dv_{\parallel}$, we obtain the continuity equation 
%\begin{equation}
%\frac{\p n}{\p t}+\nabla\cdot\lr{\bol{v}'_{\bol{E}}n}=0.
%\end{equation}
%Notice that the second term on the right-hand side of the equation for $\bol{v}_{\bol{E}}'$, which scales as $B^{-1}$, is a first order correction to $\bol{v}_{\bol{E}}$. 
We see that the polarization drift $\bol{v}_{\rm pol}$ is that average particle velocity resulting from a 
non-vanishing average acceleration $d\bol{v}_{\bol{E}}/dt$ across the magnetic field. 
%From this point on we shall identify the polarization drift $\bol{v}_{\rm pol}$ with its first order part, and write $\bol{v}_{\rm pol}=\sigma B^{-2}{\bol{B}\cp\frac{d\bol{v}_{\bol{E}}}{dt}}$   
%with $d\bol{v}_{\bol{E}}/dt=\p\bol{v}_{\bol{E}}/\p t+\bol{v}_{\bol{E}}\cdot\nabla\bol{v}_{\bol{E}}$.
%It is also worth observing that 
%\begin{equation}
%\frac{1}{\tau_c}\int_t^{t+\tau_c}\dot{\tilde{\bol{\rho}}}\,dt=-
%\sigma\frac{d}{dt}\lr{\frac{\bol{B}}{B^2}}\cp\bol{v}_{\bol{E}}+o\lr{\epsilon^2}.
%\end{equation}
%In section 4, where we compare the guiding center derivation of the gHM equation with the two-fluid approach, we will show that \eqref{vpol1} is precisely the
%expression for the polarization drift that one obtains by expanding the equation of motion of a charged particle in powers of $\sigma$. 

In the following, we shall also demand the energy $\mu B$ of cyclotron dynamics to be small, so that $\bol{v}_{\nabla}'$ becomes a higher order correction.
More precisely, denoting with $\langle\mu B\rangle=n^{-1}\int_{\mathbb{R}^3} f\mu Bd\bol{p}$ the ensemble averaged kinetic energy of cyclotron dynamics, and defining an associated temperature $T_c$ according to 
$\langle\mu B\rangle=k_BT_c$ with $k_B$ the Boltzmann constant, we demand $T_c$ to satisfy 
\begin{equation}
\frac{k_BT_c}{\frac{m}{2}\bol{v}_{\bol{E}}^2}\sim\epsilon.\label{musmall}
\end{equation}
The ordering conditions \eqref{usmall} and \eqref{musmall} can be regarded as the usual drift wave turbulence ordering requirement of cold ions. 
The guiding-center velocity thus becomes 
\begin{equation}
\dot{\bol{X}}=\bol{v}_{\bol{E}}+\bol{v}_{\rm pol}+o\lr{\epsilon^2}=\frac{\bol{b}\cp\nabla\varphi}{B}+\sigma\frac{\bol{b}\cp{\frac{d\bol{v}_{\bol{E}}}{dt}}}{B}+o\lr{\epsilon^2}.\label{XdotNgc}
\end{equation}
%There is another way to obtain \eqref{vpol1} directly from the definition of the guiding center  transformation $\bol{x}=\bol{X}+\bol{\rho}$. Indeed, from \eqref{gyro} we have
%\begin{equation}
%\dot{\bol{\rho}}=\frac{\bol{B}\cp\frac{d\lr{\dot{\bol{x}}-\bol{v}_{\bol{E}}}}{dt}}{B^2}+\frac{d}{dt}\lr{\frac{\bol{B}}{B^2}}\cp\lr{\bol{\rho}\cp\bol{B}}+o\lr{\epsilon^2}=\frac{\bol{B}\cp\frac{d\lr{\dot{\bol{x}}-\bol{v}_{\bol{E}}}}{dt}}{B^2}-\frac{d\log B^2}{dt}\bol{\rho}+o\lr{\epsilon^2},
%\end{equation}
%where all velocity contributions parallel to the magnetic field have been treated as second order corrections. 
%Hence, if on average the particle orbit remains close to the guiding center, i.e. $\frac{1}{\tau_c}\int_t^{t+\tau_c}\bol{\rho}\,dt=o\lr{\epsilon^2}$, with $\tau_c=2\pi/\omega_c$ the period of the cyclotron gyration, by averaging the equation above over a gyroperiod one obtains $\langle\dot{\bol{\rho}}\rangle=o\lr{\epsilon^2}$.
To proceed further, it is convenient to introduce the orthogonal gradient operator
\begin{equation}
\nabla_{\perp}=-\bol{b}\cp\lr{\bol{b}\times\nabla}.
\end{equation}
Although the expression \eqref{XdotNgc} is convenient to highlight the usual guiding center drift contributions separately, 
$o\lr{\epsilon^2}$ order terms must be added to $\bol{v}_{\bol{E}}+\bol{v}_{\rm pol}$ in order for the reduced (drift wave) Hamiltonian $\chi=\varphi+\frac{\sigma}{2}\bol{v}_{\bol{E}}^2$ 
arising from the expansion of the Northrop guiding center Hamiltonian $H_{Ngc}=\frac{\sigma}{2}u^2+\varphi+\mu B+\frac{\sigma}{2}\bol{v}_{\bol{E}}^2=\chi+o\lr{\epsilon}$ 
to be an exact constant of motion in the case of time-independent electromagnetic fields. 
To this end, one can verify that equation \eqref{XdotNgc}
can be equivalently written as
\begin{equation}
\dot{\bol{X}}=\dot{\bol{X}}_{dw}+o\lr{\epsilon^2},~~~~\dot{\bol{X}}_{dw}=\frac{\bol{b}\cp{\nabla\lr{\varphi+\frac{\sigma}{2}\bol{v}_{\bol{E}}^2}}}{B_{\parallel}''}-\sigma\frac{\nabla_{\perp}\varphi_t}{B^2},~~~~B_{\parallel}''=B\lr{1+\sigma\frac{\bol{b}\cdot\nabla\cp\bol{v}_{\bol{E}}}{B}},  
\end{equation}
with $\chi$ an exact integral of 
the first order term $\dot{\bol{X}}_{dw}$ when $\varphi_t=\p\varphi/\p t=0$. 

Noting that %$\bol{v}_{\bol{E}}'$ and $\bol{v}_{\rm pol}$ are 
$\dot{\bol{X}}_{dw}$ is a pure spatial function, the ensemble averaged ion velocity \eqref{avdotx} at a given position $\bol{x}=\bol{X}+{\bol{\rho}}$ is 
\begin{equation}
\langle \dot{\bol{x}}\rangle=\frac{1}{n}\int_{\mathbb{R}^3}f\lr{\dot{\bol{X}}+\dot{\bol{\rho}}}\,d\bol{p}=
\dot{\bol{X}}_{dw} 
%\bol{v}_{\bol{E}}'+\bol{v}_{\rm pol}%+\frac{\langle\mu B\rangle}{Ze}\frac{\bol{b}\times\nabla B}{B^2}
+o\lr{\epsilon^2}.\label{xdotf}
\end{equation}
%Due to \eqref{musmall}, the third term on the right-hand side of \eqref{xdotf} is now a second order term, so that 
%\begin{equation}
%\langle \dot{\bol{x}}\rangle=\bol{v}_{\bol{E}}'+\bol{v}_{\rm pol}+o\lr{\epsilon^2}.\label{xdotf2}
%\end{equation}
%where $n\lr{\bol{x},t}=\int_{\mathbb{R}^3} f\,d\bol{p}$ is the ion spatial density. 
%in the charged particle canonical phase space $\lr{\bol{p},\bol{x}}$, and assuming that collision are negligible, the Liouville equation for $f$ reads as 
%\begin{equation}
%\frac{\p f}{\p t}+
%\end{equation}
%Here, we used the fact that $\langle \dot{\bol{\rho}}\rangle=\bol{v}_{\rm pol}+o\lr{\epsilon^2}$ since the volume element $d\bol{p}$ can be mapped to the guiding center variables $\mu$, $u$, and $\zeta$, and thus the average above includes integration with respect to the gyrophase $\zeta$.
Next, consider the density $n_e\lr{\bol{x},t}$ of the electron component. 
%Notice that $n_e$ is being evaluated at the electron position $\bol{x}$. 
We assume that $n_e$ follows a Boltzmann distribution with 
temperature $T_e$, i.e.
\begin{equation}
n_e=A_e\lr{\bol{x}}\exp\lrc{\lambda{\varphi\lr{\bol{x},t}}},~~~~\lambda=\frac{e}{k_BT_e}%,~~~~\sigma_e=\frac{m}{e}
\label{ne}
\end{equation}
where $A_e\lr{\bol{x}}$ is a spatial function. 
%Notice that the kinetic energy associated with electron $\bol{E}\cp\bol{B}$ motion appears explicitly within the exponential, a fact that will later ensure conservation of total energy for the drift wave turbulence system.  
If we further demand the ion-electron plasma to be quasi-neutral, 
%and neglect the oscillatory gyrophase dependent part $\tilde{\bol{\rho}}$ of the gyroradius, 
we have the following condition:  
\begin{equation}
Zn\lr{\bol{x},t}=n_e\lr{\bol{x},t}.\label{nne}
\end{equation}
Then, %neglecting second order terms,  %denoting with $\nabla_{\bol{x}}$ the gradient operator with respect to $\bol{x}=\bol{X}+\bar{\bol{\rho}}$, 
the continuity equation for the ion density reads as 
\begin{equation}
Z\lrs{\frac{\p n}{\p t}+\nabla\cdot\lr{\langle\dot{\bol{x}}\rangle n}}=\frac{\p n_e}{\p t}+\nabla\cdot \lr{\dot{\bol{X}}_{dw}n_e}+o\lr{\epsilon^2}=0,\label{gHM0}
\end{equation}
where we used equation \eqref{xdotf}. 
Substituting equation \eqref{ne}, equation \eqref{gHM0} can be rearranged as
\begin{equation}
\begin{split}
\lambda A_e\frac{\p\varphi}{\p t}%\lr{\varphi+\frac{\sigma}{2}\bol{v}_{\bol{E}}^2}
=&-\lambda\nabla{\varphi}\cdot{A_e\dot{\bol{X}}_{dw}}-\nabla\cdot\lr{A_e\dot{\bol{X}}_{dw}}+o\lr{\epsilon^2}\\=&\lambda\sigma A_e\nabla{\varphi}\cdot\lr{ \frac{\nabla_{\perp}\varphi_t}{B^2}+\frac{\bol{b}\cp\nabla\bol{v}_{\bol{E}}^2}{2B_{\parallel}''}}-\nabla\cdot\lrs{A_e\frac{\bol{b}\cp\nabla\lr{\varphi+\frac{\sigma}{2}\bol{v}_{\bol{E}}^2}}{B_{\parallel}''}}+\sigma\nabla\cdot\lr{A_e\frac{\nabla_{\perp}\varphi_t}{B^2}}+o\lr{\epsilon^2}.\label{GHMgc1}
\end{split}
\end{equation}
Next, we demand the electron component to be hot compared to the ion component, i.e.
\begin{equation}
\frac{\frac{m}{2}\bol{v}_{\bol{E}}^2}{k_BT_e}\sim\epsilon^2.
\end{equation}
Since $\lambda=e/k_BT_e$, it follows that $\lambda\varphi\sim\epsilon$, while the first  term on the right-hand side %and the first term on the right-hand side 
of equation \eqref{GHMgc1} scales as $\epsilon^2$. Equation \eqref{GHMgc1} thus reduces to
\begin{equation}
\begin{split}
\frac{\p}{\p t}&\lrs{\lambda A_e\varphi-\sigma\nabla\cdot\lr{A_e\frac{\nabla_{\perp}\varphi}{B^2}}}=-\nabla\cdot\lrs{A_e\frac{\bol{b}\cp\nabla\lr{\varphi+\frac{\sigma}{2}\bol{v}_{\bol{E}}^2}}{B_{\parallel}''}}+o\lr{\epsilon^2}.
%+\sigma{\lr{\bol{v}_{\bol{E}}\cdot\nabla}\lr{\frac{\bol{B}}{B^2}}}\cp\bol{v}_{\bol{E}}
\end{split}\label{GHMgc2}
\end{equation}
\begin{table}
\begin{center}
\begin{tabular}{c c c c c c} 
 \hline
 \hline
 Order & Dimensionless & Fields & Distances & Rates & Velocities\\  
 \hline 
 $\epsilon^{-1}$ & & $\bol{B},\bol{E}_{\perp}$& & $\omega_c$ &\\ 
 $1$ &  & $A_e$ & $L$ & $\tau^{-1}_d,\bol{v}_{\bol{E}}/L$ & $\bol{v}_{\bol{E}}$\\
 $\epsilon$ &  $\lambda\varphi,\rho/L,\lr{\omega_c\tau_d}^{-1}, k_BT_c/\frac{m}{2}\bol{v}_{\bol{E}}^2$ &  & $\rho$ & $\bol{v}_{\rm pol}/L$ & $\bol{v}_{\rm pol}$\\
 $\epsilon^2$ &$\frac{m}{2}\bol{v}_{\bol{E}}^2/k_BT_e,\tau_d/\tau_b$ & $E_{\parallel}'$ &  & $\bol{v}_{\nabla}/L,u/L,\tau_b^{-1}$ & $\bol{v}_{\nabla},u$ \\
%$\epsilon^3$& $\sigma\abs{\nabla\bol{v}_{\bol{E}}^2\cdot\nabla\cp\lr{A_e\frac{\bol{B}}{B^2}}}/A_e\omega_c$ & & & &\\
 $\epsilon^5$ & & & & $\bol{v}_{\kappa}/L$ & $\bol{v}_{\kappa}$ \\
 \hline
 \hline
\end{tabular}
\caption{\label{tab2} Drift wave turbulence ordering used  for the derivation of the GHM equation within the guiding-center framework.}
\end{center}
\end{table}
Since this equation is correct up to first order in $\epsilon$, we can add terms scaling as $\epsilon^2$ that will be useful later to obtain exact conservation laws: 
\begin{equation}
\begin{split}
\frac{\p}{\p t}&\lrc{\lambda A_e\lr{\varphi+\frac{\sigma}{2}\bol{v}_{\bol{E}}^2}-\sigma\nabla\cdot\lrs{A_e\frac{\nabla_{\perp}\lr{\varphi+\frac{\sigma}{2}\bol{v}_{\bol{E}}^2}}{B^2}}}=-\nabla\cdot\lrs{A_e\frac{\bol{b}\cp\nabla\lr{\varphi+\frac{\sigma}{2}\bol{v}_{\bol{E}}^2}}{B_{\parallel}''}}+o\lr{\epsilon^2}.
%+\sigma{\lr{\bol{v}_{\bol{E}}\cdot\nabla}\lr{\frac{\bol{B}}{B^2}}}\cp\bol{v}_{\bol{E}}
\end{split}\label{GHMgc3}
\end{equation} 
Recalling that the Northrop guiding center Hamiltonian reduces to  
\begin{equation}
\chi=\varphi+\frac{\sigma}{2}\bol{v}_{\bol{E}}^2, 
\end{equation}
in the drift wave turbulence ordering, 
and that $B_{\parallel}''=B\lr{1+\sigma\frac{\bol{b}\cdot\nabla\cp\bol{v}_{\bol{E}}}{B}}$ 
%We thus arrive at the following closed equation for the potential  $\varphi$,
we thus arrive at the equation 
\begin{equation}
\begin{split}
\frac{\p}{\p t}\lrs{\lambda A_e\chi-\sigma\nabla\cdot\lr{A_e\frac{\nabla_{\perp}\chi}{B^2}}}&=-\nabla\cdot\lr{A_e\frac{\bol{b}\cp\nabla\chi}{B_{\parallel}''}}+o\lr{\epsilon^2}\\&=\nabla\cdot\lrc{A_e\lrs{\sigma\frac{\bol{b}\cdot\nabla\cp\lr{\frac{\bol{b}\cp\nabla\chi}{B}}}{B}-1}\frac{\bol{b}\cp\nabla\chi}{B}}+o\lr{\epsilon^2}. 
%,~~~~\chi=\varphi+\frac{\sigma}{2}\bol{v}_{\bol{E}}^2.
%\frac{\p}{\p t}&\lrs{\lambda A_e\varphi-\sigma\nabla\cdot\lr{A_e\frac{\nabla_{\perp}\varphi}{B^2}}}=\nabla\cdot\lrs{A_e{\lr{\sigma\frac{\bol{B}\cdot\nabla\cp\bol{v}_{\bol{E}}}{B^2}-1}\bol{v}_{\bol{E}}}}.
%+\sigma{\lr{\bol{v}_{\bol{E}}\cdot\nabla}\lr{\frac{\bol{B}}{B^2}}}\cp\bol{v}_{\bol{E}}
\end{split}\label{GHMgc4}
\end{equation}
Notice that $\chi=\varphi+o\lr{1}$. 
If we further define the vector field
\begin{equation}
\bol{v}_{\bol{E}}^{\chi}=\frac{\bol{b}\cp\nabla\chi}{B}=\bol{v}_{\bol{E}}+\sigma\frac{\bol{b}\cp\nabla\bol{v}_{\bol{E}}^2}{2B},%~~~~\dot{\bol{X}}_{dw}=\bol{v}_{\bol{E}}^{\chi}-\sigma\frac{\nabla_{\perp}\varphi_t}{B^2},
\end{equation}
equation \eqref{GHMgc4} 
gives the following closed equation for the variable $\chi$, 
\begin{equation}
\frac{\p}{\p t}\lrs{\lambda A_e\chi-\sigma\nabla\cdot\lr{A_e\frac{\nabla_{\perp}\chi}{B^2}}}=\nabla\cdot\lrs{A_e\lr{\sigma\frac{\bol{B}\cdot\nabla\cp\bol{v}_{\bol{E}}^{\chi}}{B^2}-1}\bol{v}_{\bol{E}}^{\chi}}.\label{GHMgc5}
\end{equation}
In the following, we shall refer to equation \eqref{GHMgc5} as the guiding center generalized Hasegawa-Mima equation (GHM) to distinguish it from the gHM equation \eqref{eq1} derived in \cite{gHM} from a two-fluid plasma model. Nonetheless, notice that these two equations share the same mathematical form, with the variable $\chi$ in \eqref{GHMgc5} replacing $\varphi$ in \eqref{eq1}.  
The relationship between the ordering conditions used to construct these two models will be discusses in  section 4.  
Here, we observe that in \eqref{GHMgc5} %there %are two terms, $\lambda A_e\varphi_t$ and  
%is one term, 
the term 
$\nabla\cdot\lr{A_e\bol{v}_{\bol{E}}^{\chi}}=\nabla\chi\cdot\nabla\cp\lr{A_e\bol{B}/B^{2}}$ %, which scales as $\epsilon^0\sim 1$. 
must scale as $\sim\epsilon$ to be consistent with the other terms in the equation. Hence, either $\nabla\cp\lr{A_e\bol{B}/B^{2}}$ scales as $\sim\epsilon^2$, or the effective electric field  $-\nabla\chi\sim\epsilon^{-1}$ is mostly orthogonal to  the vector field $\nabla\cp\lr{A_e\bol{B}/B^{2}}\sim\epsilon$. 
We stress however that the behavior of the term $\nabla\cdot\lr{A_e\bol{v}_{\bol{E}}^{\chi}}$ is a consequence of the ordering used to obtain equation \eqref{GHMgc5}, 
and not an ordering condition required to arrive at \eqref{GHMgc5}. 

%, and therefore they must balance each other. This is consistent with the assumption that $\lambda\varphi\sim\epsilon$. Indeed, 
%if $\lambda\varphi$ contains fast oscillations, its time derivative can be a lower order contribution. For example, if $\lambda\varphi\propto\epsilon\sin\lr{t/\epsilon}\sim \epsilon$ one has $\lambda\varphi_t\propto\epsilon d\sin\lr{t/\epsilon}/dt\sim 1$. 
The ordering used to derive the GHM equation \eqref{GHMgc5} is summarized in table \ref{tab2}. 
Finally, one can verify that equation \eqref{GHMgc5} reduces to the standard HM equation 
%by the change of variables $\chi\rightarrow\varphi$,  
\begin{equation}
\frac{\p}{\p t}\lr{\lambda\varphi-\frac{\sigma}{B_0^2}\Delta_{\lr{x,y}}\varphi}=\frac{\sigma}{B_0^3}\lrs{\varphi,\Delta_{\lr{x,y}}\varphi}_{\lr{x,y}}+\frac{\beta}{B_0}\varphi_y,\label{HM}
\end{equation}
when $\bol{B}=B_0\nabla z$, $\log A_{e}=\log A_{e0}+\beta x$, $B_0,A_{e0},\beta\in\mathbb{R}$, $\beta L\sim\epsilon$. Here, $\lrs{f,g}_{\lr{x,y}}=f_xg_y-f_yg_x$,  $\Delta_{\lr{x,y}}=\p_{x}^2+\p_{y}^2$, and lower indexes denote partial derivatives, for example $f_x=\p f/\p x$.

\section{Conservation laws}
In this section, we show that the derived GHM equation \eqref{GHMgc5} preserves both mass and energy. 
Furthermore, we identify a third invariant  associated with the vorticity of the flow in a more general form than the one obtained in \cite{gHM}, and discuss its relationship with the generalized enstrophy encountered in the  standard HM equation. 
%From this point on we shall use the variable $\varphi$ in place of $\chi$   
%to simplify the notation while keeping in mind that 

Since the GHM equation \eqref{GHMgc5} and the gHM equation \eqref{eq1} 
share the same mathematical structure, we already know that the 
invariants of the GHM equation \eqref{GHMgc5} can be obtained by replacing $\varphi$ 
with $\chi$ in the expressions of the invariants of the gHM equation \eqref{eq1}. 
It is however useful to recall the physical origin of these quantities. 
%\begin{equation}
%\lambda A_e\chi\,dV
%\end{equation}
%\begin{equation}
%\frac{1}{2}A_e\lr{\lambda\chi^2+\sigma\frac{\abs{\nabla_{\perp}\chi}^2}{B^2}}\,dV
%\end{equation}
First observe that the the total ion mass can be written as
\begin{equation}
\mc{M}_{\Omega}=\frac{m}{Z}\int_{\Omega} A_ee^{\lambda\varphi}\,d\bol{x}.
\end{equation}
Since $\lambda\varphi\sim\epsilon$, we may expand the exponential in powers of $\lambda\varphi$ according to $e^{\lambda\varphi}=1+\lambda\chi+o\lr{\epsilon^2}$ and consider the conservation of the first order term,
\begin{equation}
M_{\Omega}=\frac{m}{Z}\int_{\Omega}A_e\lr{1+\lambda\chi}d\bol{x}.\label{M}
\end{equation}
Using \eqref{GHMgc5} we have
\begin{equation}
\frac{dM_{\Omega}}{dt}=\frac{m}{Z}\int_{\p\Omega}A_e\lrs{\sigma{\frac{\nabla_{\perp}\chi_t}{B^2}}-{\lr{1-\sigma\frac{\bol{B}\cdot\nabla\cp\bol{v}_{\bol{E}}^{\chi}}{B^2}}\bol{v}_{\bol{E}}^{\chi}}}\cdot\bol{n}\,dS=-\frac{m}{Z}\int_{\p\Omega}A_e\dot{\bol{X}}_{dw}'\cdot\bol{n}\,dS,\label{eq37}
%+\sigma\lr{\bol{v}_{\bol{E}}\cdot\nabla}\lr{\frac{\bol{B}}{B^2}}\cp\bol{v}_{\bol{E}}
\end{equation}
where $\bol{n}$ denotes the unit outward normal to the bounding surface $\p\Omega$, $dS$ the surface element on $\p\Omega$, and we defined
\begin{equation}
\dot{\bol{X}}_{dw}'={\lr{1-\sigma\frac{\bol{B}\cdot\nabla\cp\bol{v}_{\bol{E}}^{\chi}}{B^2}}\bol{v}_{\bol{E}}^{\chi}}-\sigma{\frac{\nabla_{\perp}\chi_t}{B^2}}=\dot{\bol{X}}_{dw}+o\lr{\epsilon^2}.
\end{equation}
The boundary integral \eqref{eq37} vanishes under suitable boundary conditions, such as $A_e=0$ on $\p\Omega$ or $\dot{\bol{X}}_{dw}'\cdot\bol{n}=0$ on $\p\Omega$. 

Next, observe that the leading order ion Hamiltonian is given by 
\begin{equation}
H=\frac{m}{2}{\bol{v}}^2+Ze\varphi=Ze\varphi+\frac{m}{2}\bol{v}_{\bol{E}}^2+o\lr{\epsilon}.
\end{equation}
Therefore, at leading order the total ion energy satifies
%\begin{equation}
%\mc{H}_{\Omega}=\frac{1}{Z}\int_{\Omega}A_e \lr{1+\lambda\varphi}\lr{\varphi+\frac{\sigma}{2}\bol{v}_{\bol{E}}^2}\,d\bol{p}d\bol{x}
%\end{equation}
\begin{equation}
\mc{H}_{\Omega}=Ze\int_{\Omega\times\mathbb{R}^3} f\lr{\varphi+\frac{\sigma}{2}\bol{v}_{\bol{E}}^2}\,d\bol{p}d\bol{x}+o\lr{\epsilon}=e\int_{\Omega}A_ee^{\lambda\varphi}\lr{\varphi+\frac{\sigma}{2}\bol{v}_{\bol{E}}^2}\,d\bol{x}+o\lr{\epsilon}.
\end{equation}
Dividing this expression by $k_BT_e$ we obtain
\begin{equation}
\frac{\mc{H}_{\Omega}}{k_BT_e}=\int_{\Omega}A_e\lr{1+\lambda\varphi+\frac{1}{2}\lambda^2\varphi^2}\lr{\lambda\varphi+\frac{\sigma}{2}\lambda\bol{v}_{\bol{E}}^2}\,d\bol{x}+o\lr{\epsilon^4}=\int_{\Omega}A_e\lr{\lambda\varphi+\lambda^2\varphi^2+\frac{\sigma}{2}\lambda\bol{v}_{\bol{E}}^2}\,d\bol{x}+o\lr{\epsilon^4}.
\end{equation}
It follows that
\begin{equation}
\frac{\mc{H}_{\Omega}}{k_BT_e}-\frac{Z}{m}{\mc{M}_{\Omega}}=\frac{1}{2}\int_{\Omega}A_e\lr{\lambda^2\varphi^2+\lambda\sigma\bol{v}_{\bol{E}}^2-2}\,d\bol{x}+o\lr{\epsilon^3}=\frac{1}{2}\int_{\Omega}A_e\lr{\lambda^2\chi^2+\lambda\sigma\bol{v}_{\bol{E}}^{\chi2}-2}\,d\bol{x}+o\lr{\epsilon^3}.
\end{equation}
Since $A_e$ is a spatial function, we thus expect the 
GHM energy
\begin{equation}
H_{\Omega}=\frac{1}{2}\int_{\Omega}A_e\lr{\lambda\chi^2+\sigma\frac{\abs{\nabla_{\perp}\chi}^2}{B^2}}\,d\bol{x},\label{gHMH}
\end{equation}
to be a constant of motion. 
From equation \eqref{GHMgc5}, one can verify that
\begin{equation}
\begin{split}
\frac{dH_{\Omega}}{dt}=&
\int_{\Omega}\chi\frac{\p}{\p t}
\lrs{\lambda A_e\chi-
\sigma\nabla\cdot\lr{A_e\frac{\nabla_{\perp}\chi}{B^2}}}\,d\bol{x}+\sigma\int_{\p\Omega}A_e\chi\frac{\nabla_{\perp}\chi_t
}{B^2}\cdot\bol{n}\,dS\\=&
\int_{\p\Omega}A_e\chi\lrs{\sigma\frac{\nabla_{\perp}\chi_t}{B^2}-\lr{1-\sigma\frac{\bol{B}\cdot\nabla\times\bol{v}_{\bol{E}}^{\chi}}{B^2}}\bol{v}_{\bol{E}}^{\chi}}\cdot\bol{n}\,dS\\=&-\int_{\p\Omega}A_e\chi\dot{\bol{X}}_{dw}'\cdot\bol{n}\,dS.
\end{split}
\end{equation}
Again, this boundary integral vanishes under suitable boundary conditions, such as $A_e=0$ on $\p\Omega$, $\chi=0$ on $\p\Omega$, or $\dot{\bol{X}}_{dw}'\cdot\bol{n}=0$ on $\p\Omega$. 
\begin{table}
\begin{center}
\begin{tabular}{c c c c c c} 
 \hline
 \hline
 Invariant & Expression & Field Conditions & Boundary Conditions\\  
 \hline 
 Mass $M_{\Omega}$ & $\int_{\Omega}A_e\lr{1+\lambda\chi}\,d\bol{x}$ & none &  $A_e\dot{\bol{X}}_{dw}'\cdot\bol{n}=0$ &\\
 Energy $H_{\Omega}$ & $\frac{1}{2}\int_{\Omega}A_e\lr{\lambda\chi^2+\sigma\frac{\abs{\nabla_{\perp}\chi}^2}{B^2}}\,d\bol{x}$ & none & $A_e\chi\dot{\bol{X}}_{dw}'\cdot\bol{n}=0$&\\
% Vorticity $G_{\Omega}$ & $-\sigma\int_{\Omega}\nabla\cdot\lr{A_e\frac{\nabla_{\perp}\chi}{B^2}}\,d\bol{x}$ & none & $A_e\bol{v}_{\bol{E}}\cdot\bol{n}=0$ &\\
Enstrophy $W_{\Omega}$ & $\int_{\Omega}A_ew\lr{\lambda\chi-\frac{\sigma}{A_e}\nabla\cdot\lr{A_e\frac{\nabla_{\perp}\chi}{B^2}}}\,d\bol{x}$ & $\nabla\cp\lr{A_e\frac{\bol{B}}{B^2}}=\bol{0}$ & $wA_e\bol{v}_{\bol{E}}^{\chi}\cdot\bol{n}=0$ &\\
 \hline
 \hline
\end{tabular}
\caption{\label{tab3} Invariants of the GHM equation \eqref{GHMgc5}.}
\end{center}
\end{table}

An additional invariant, associated with the vorticity $\nabla\cp\bol{v}_{\bol{E}}^{\chi}$, 
 exists when the magnetic field and the electron spatial density satisfy the condition $\nabla\cp\lr{A_e\bol{B}/B^2}=\bol{0}$. 
%when $A_e\bol{v}_{\bol{E}}\cdot\bol{n}=0$ on $\p\Omega$. 
To see this, define the quantity 
\begin{equation}
\omega=\nabla\cdot\lr{A_e\frac{\nabla_{\perp}\chi}{B^2}}=A_e\frac{\bol{B}\cdot\nabla\cp\bol{v}_{\bol{E}}^{\chi}}{B^2}+\nabla\chi\cdot\frac{\bol{B}}{B^2}\cp\lrs{\nabla\cp\lr{A_e\frac{\bol{B}}{B^2}}}.
\end{equation}
%Hence, from \eqref{gHM2} it follows that the functional
%\begin{equation}
%G_{\Omega}=-\sigma\int\omega\,d\bol{x},
%\end{equation}
%satisfies
%\begin{equation}
%\frac{dG_{\Omega}}{dt}=\int_{\Omega}\frac{\p}{\p t}\lr{\lambda A_e\varphi-\sigma\omega}d\bol{x}-\frac{dM_{\Omega}}{dt}=-\int_{\p\Omega}A_e\lr{1-\sigma\frac{\bol{B}\cdot\nabla\cp\bol{v}_{\bol{E}}}{B^2}}\bol{v}_{\bol{E}}\cdot\bol{n}\,dS.
%\end{equation}
%When the aforementioned boundary condition  $A_e\bol{v}_{\bol{E}}\cdot\bol{n}=0$ on $\p\Omega$ holds, 
%the quantity $G_{\Omega}$ is therefore a constant of motion.
Next, observe that whenever
\begin{equation}
\frac{\bol{B}}{B^2}\cp\lrs{\nabla\cp\lr{A_e\frac{\bol{B}}{B^2}}}=\bol{0},\label{Bel}
\end{equation}
which implies that $A_e\bol{B}/B^2$
is a Beltrami field, the following identity holds 
\begin{equation}
\omega=\nabla\cdot\lr{A_e\frac{\nabla_{\perp}\chi}{B^2}}=A_e\frac{\bol{B}\cdot\nabla\cp\bol{v}_{\bol{E}}^{\chi}}{B^2}.
\end{equation}
The derived GHM equation \eqref{GHMgc5} can thus be written in the form
\begin{equation}
\frac{\p}{\p t}\lrs{\lambda A_e\chi-\sigma\omega}=-\nabla\cdot\lrs{\lr{1-\sigma\frac{\omega}{A_e}}A_e\bol{v}_{\bol{E}}^{\chi}}.
\end{equation}
On the other hand, functionals of the form
\begin{equation}
W_{\Omega}=\int_{\Omega}A_ew\lr{\lambda\chi-\sigma\frac{\omega}{A_e}}\,d\bol{x},\label{W}
\end{equation}
where $w\lr{\lambda\chi-\sigma\omega/A_e}$ is any  function of $\lambda\chi-\sigma\omega/A_e$, satisfy
\begin{equation}
\frac{dW_{\Omega}}{dt}=\int_{\Omega}w'\frac{\p}{\p t}\lr{\lambda A_e\chi-\sigma\omega}\,d\bol{x}=-\int_{\Omega}w'\lrs{\nabla\lr{\lambda\chi-\sigma\frac{\omega}{A_e}}\cdot A_e\bol{v}_{\bol{E}}^{\chi}+\lr{1-\sigma\frac{\omega}{A_e}}\nabla\cdot\lr{A_e\bol{v}_{\bol{E}}^{\chi}}}\,d\bol{x},
\end{equation}
with $w'=d w/d\lr{\lambda\chi-\sigma\omega/A_e}$. 
Now observe that
\begin{equation}
\nabla\cdot\lr{A_e\bol{v}_{\bol{E}}^{\chi}}=\nabla\chi\cdot\nabla\cp\lr{A_e\frac{\bol{B}}{B^2}}.
\end{equation}
Hence, if we further demand that
\begin{equation}
\nabla\cp\lr{A_e\frac{\bol{B}}{B^2}}=\bol{0},\label{norm}
\end{equation}
which is a special case of \eqref{Bel}, we find that
\begin{equation}
\frac{dW_{\Omega}}{dt}=-\int_{\p\Omega} wA_e\bol{v}_{\bol{E}}^{\chi}\cdot\bol{n}\,dS.
\end{equation}
This boundary integral vanishes whenever $wA_e\bol{v}_{\bol{E}}^{\chi}\cdot\bol{n}=0$ on $\p\Omega$. 
The quantity $W_{\Omega}$ can be identified with the generalized enstrophy encountered in the standard Hasegawa-Mima equation if the boundary condition above is satisfied through $A_e$, i.e. $A_e=0$ on $\p\Omega$. Indeed, choosing $w=\lr{\lambda\chi-\sigma\omega/A_e}^2$ and integrating by parts gives
\begin{equation}
W_{\Omega}=2\lambda H_{\Omega}+\sigma\int_{\Omega}\lrc{\lambda A_e\frac{\abs{\nabla_{\perp}\chi}^2}{B^2}+\sigma A_e^{-1}\lrs{\nabla\cdot\lr{A_e\frac{\nabla_{\perp}\chi}{B^2}}}^2}\,d\bol{x}.
\end{equation}
In the following, we shall refer to $W_{\Omega}$ as the generalized enstrophy. 
It is worth observing that the condition \eqref{norm} 
implies (Poincar\'e lemma) that the magnetic field locally defines the normal direction of a surface $C={\rm constant}$, i.e. 
$\bol{B}\propto \nabla C$ for some appropriate function $C$ and sufficiently small neighborhood $U\subseteq\Omega$. 

The invariants of the GHM equation are summarized in table \ref{tab3}.

%Recalling that in the guiding-center ordering of table \ref{tab1}
%the gyroradius is a first order term, $\bar{\bol{\rho}}\sim \epsilon L$, 
%equation \eqref{nne} can be expanded as follows:
%\begin{equation}
%Zn=n_e+\nabla n_e\cdot\frac{\bol{b}\cp\bol{v}_{\bol{E}}}{\omega_c}+o\lr{\epsilon^2},
%\end{equation}
%where we used equation \eqref{vpol} and all quantities are now evaluated at the ion guiding-center position $\bol{X}$.

\section{Relationship with the two-fluid model gHM}
%effective potential energy/force that does not appear in fluid model>why? Everything seems to be related to how the polarization displacement is evaluated ... 
In this section, we discuss the relationship between the ordering conditions used to derive the gHM equation \eqref{eq1} from a two-fluid model in \cite{gHM}, and those used above to derive the GHM equation \eqref{GHMgc5} within the kinetic framework of guiding-center dynamics. 
Recall that the two equations possess the same mathematical form, but the 
respective orderings are slightly different. 
\begin{table}
\begin{center}
\begin{tabular}{c c c c c c} 
 \hline
 \hline
 Order & Dimensionless & Fields & Distances & Rates & Velocities\\  
 \hline 
 $1$ & & $\bol{B},A_e$& $L$ & $\omega_c$ &\\ 
 $\epsilon$ & $\lambda\varphi, %\frac{\p\log\abs{\bol{v}_{\bol{E}}}}{\p \omega_ct},
 %\frac{\p\log\abs{\bol{v}_{\rm pol}}}{\p \omega_ct},\frac{\p\log\abs{v_{\parallel}}}{\p \omega_ct}
 \omega_c^{-1}\p_t$ & $\bol{E}_{\perp}$ & & $\tau^{-1}_d,\bol{v}_{\bol{E}}/L$ & $\bol{v}_{\bol{E}}$\\
 $\epsilon^2$ &  $\tau_d/\tau_b$ &  &  & $\bol{v}_{\rm pol}/L$ & $\bol{v}_{\rm pol}$\\
 $\epsilon^3$ &$\sigma\abs{\nabla\bol{v}_{\bol{E}}^2\cdot\nabla\cp\lr{A_e\frac{\bol{B}}{B^2}}}/A_e\omega_c$ & $E_{\parallel},P$ &  & $\tau_b^{-1},v_{\parallel}/L$ & $v_{\parallel}$ \\
  \hline
 \hline
\end{tabular}
\caption{\label{tab4} Drift wave turbulence ordering used  for the derivation of the gHM equation from a two-fluid model in \cite{gHM}. Here, $P$ denotes the ion fluid pressure and $v_{\parallel}$ the ion fluid velocity along $\bol{B}$. }
\end{center}
\end{table}
\begin{table}
\begin{center}
\begin{tabular}{c c c c c c} 
 \hline
 \hline
 Order & Dimensionless & Fields & Distances & Rates & Velocities\\  
 \hline 
 $1$ & & $\bol{B},A_e$& $L$ & $\omega_c$ &\\ 
 $\epsilon$ & $\lambda\varphi,\omega_c^{-1}\p_t,L\nabla\log B,L\nabla \log A_e$ & $\bol{E}_{\perp}$ & & $\tau^{-1}_d,\bol{v}_{\bol{E}}/L$ & $\bol{v}_{\bol{E}}$\\
 $\epsilon^2$ & $\tau_d/\tau_b$  &  &  & $\bol{v}_{\rm pol}/L$ & $\bol{v}_{\rm pol}$\\
 $\epsilon^3$ & & $E_{\parallel},P$ &  & $\tau_b^{-1},v_{\parallel}/L$ & $v_{\parallel}$ \\
  \hline
 \hline
\end{tabular}
\caption{\label{tab5} Drift wave turbulence ordering used  for the derivation of the standard HM equation in a straight homogeneous magnetic field from a two-fluid model. Here, $P$ denotes the ion fluid pressure and $v_{\parallel}$ the ion fluid velocity along $\bol{B}$.}
\end{center}
\end{table}

Table \ref{tab4} summarizes the drift wave turbulence ordering used to derive the gHM equation from a two-fluid model in \cite{gHM}, while the standard drift wave turbulence ordering for the HM equation \eqref{HM} is given in table \ref{tab5}. 
Here, we observe that the standard HM ordering of table \ref{tab5} is stricter than the two-fluid gHM ordering of table \ref{tab4}. 
In particular, the HM conditions on the spatial changes in $B$ and $A_e$, $L\nabla\log B\sim L\nabla\log A_e\sim\epsilon$ are relaxed through the gHM condition
\begin{equation}
\frac{\sigma}{A_e\omega_c}\abs{\nabla\bol{v}_{\bol{E}}^2\cdot\nabla\cp\lr{A_e\frac{\bol{B}}{B^2}}}\sim\epsilon^3.\label{gHMB}
\end{equation}
%while the HM condition $\omega_c^{-1}\p_t\sim\epsilon$ 
%only applies to the velocity components $\bol{v}_{\bol{E}}$, $\bol{v}_{\rm pol}$, and $v_{\parallel}$ in the gHM case ($v_{\parallel}$ is the ion fluid velocity parallel to $\bol{B}$). 

If we now compare the GHM ordering of table \ref{tab2} with the gHM ordering of table \ref{tab4}, one first notices that $\bol{B}$ scales as $\epsilon^{-1}$ in the GHM case, while it is treated as a $\epsilon^0\sim 1$ term in the gHM ordering. This difference does not change the order of the ratio ${v}_{\bol{E}}/{v}_{\rm pol}$, and it is therefore not essential (in fact, all dimensionless ratios have the same order in both orderings; compare the column `dimensionless' in \ref{tab2} and \ref{tab4} with the fluid pressure $P$ playing the role of the cyclotron temperature $T_c$). 
The key difference is the absence of 
any ordering requirement on derivatives of $\bol{B}$ and $A_e$ in the GHM ordering, and, in particular, the absence of the condition \eqref{gHMB} found in the gHM case. This implies that the GHM model is free from any requirements on the geometry of the magnetic field $\bol{B}$ or the spatial density $A_e$, and turbulence may develop  over spatial scales $k_{\perp}^{-1}$ comparable to $1/\abs{\nabla\log B}$ and $1/\abs{\nabla\log A_e}$.  
This difference does not originate from a discrepancy between two-fluid theory and
guiding center theory. Indeed, the ordering condition \eqref{gHMB} used in \cite{gHM} 
to enforce conservation of energy for the gHM equation \eqref{eq1} can be removed by
adding higher order terms to \eqref{eq1} 
through the same procedure used to arrive at the GHM equation \eqref{GHMgc5}.
Hence, one can obtain the same equation \eqref{GHMgc5} from two fluid theory as well.

\section{Concluding remarks}
In this study, we derived the GHM equation \eqref{GHMgc5} describing drift wave turbulence in curved magnetic fields within the framework of guiding center dynamics. 
This equation exhibits the same mathematical structure of the gHM equation \eqref{eq1} derived in \cite{gHM} from a two-fluid model of an ion-electron plasma: the GHM equation can be obtained from the gHM equation by replacing the electrostatic potential $\varphi$ with the reduced Northrop guiding center Hamiltonian $\chi=\varphi+\frac{\sigma}{2}\bol{v}_{\bol{E}}^2$. 
However, the ordering (table \ref{tab2}) used to obtain the GHM equation is weaker 
than the ordering  (table \ref{tab4}) used in \cite{gHM} to derive the gHM equation.
In particular, while the gHM equation relies on the ordering condition \eqref{gHMB} to enforce conservation of energy, the GHM ordering does not involve ordering conditions on spatial derivatives of the magnetic field $\bol{B}$ or the electron spatial density $A_e$. 
%This is in contrast with the two-fluid model gHM equation of \cite{gHM} where the ordering condition \eqref{gHMB} is assumed to enforce conservation of energy (see also table \ref{tab4}). 
As explained in section 4, the GHM equation can also be obtained from two-fluid theory as well 
by adding higher-order terms to the gHM equation. 
%The relationship between the GHM equation \eqref{GHMgc5} and the gHM equation \eqref{eq1}, which has been discussed in section 4
These results imply that the GHM equation is appropriate to describe 
drift wave turbulence in general magnetic fields and 
in the presence of strong inhomogeneities in the electron spatial density, 
including the case in which the scale of spatial change 
in the magnetic field and the electron spatial density 
is comparable to that of the turbulent electric field. 
This setting is expected to be physically relevant in strongly curved and  inhomogeneous magnetic fields, such as dipole magnetic fields or 
the confining magnetic field of a stellarator. 
Finally, we remark that for practical purposes (e.g. numerical implementation), 
the solution $\chi$ of the GHM equation can be used to
approximate $\varphi\approx\chi$ since both $\chi$ and $\varphi=\chi+o\lr{1}$ scale as $\epsilon^{-1}$.

\section*{Statements and declarations}

\subsection*{Data availability}
Data sharing not applicable to this article as no datasets were generated or analysed during the current study.

\subsection*{Funding}
The research of NS was partially supported by JSPS KAKENHI Grant No. 21K13851.
and 22H04936.

\subsection*{Competing interests} 
The authors have no competing interests to declare that are relevant to the content of this article.

%\section*{Data Availability}

%The data that support the findings of this study are available from
%the corresponding author upon reasonable request.	

\end{document}